# Research directions for kidney stone disease

Vincent Blay[a], Felix Grases[b]

[a] Department of Microbiology and Environmental Toxicology, University of California, Santa Cruz, Santa Cruz, CA 95064, USA. vroger@ucsc.edu

[b] Laboratory of Renal Lithiasis Research, University Institute of Health Sciences Research (IUNICS-IdISBa), University of Balearic Islands, 07122 Palma de Mallorca, Spain. fgrases@uib.es

ORCID: 0000-0001-9602-2375 (V.B.), 0000-0001-5636-9139 (F.G.).

Kidney stone disease poses a major burden to patients and healthcare systems around the world [1]. The formation of kidney stones may occur over months or years, but many patients are diagnosed at a late stage, suffer excruciating pain, and require surgical intervention to physically remove the stones. The prevalence of kidney stones has increased during recent decades to over 10% in many developed countries [1, 2], suggesting a link with environmental and behavioral factors. Recurrence rates are also high [3]. In terms of their impact and scale, kidney stones are an ongoing pandemic. The causes and mechanisms of kidney stone formation are diverse and often unknown, resulting in varied compositions and different anatomical locations being affected [1]. A better understanding of these processes could enable earlier diagnoses through more sensitive and scalable biomarkers, as well as more effective preventives and therapeutics.

The characterization of calculi from patients is becoming more common, as it can provide key information about their formation or etiology and help guide treatment [4]. The study should cover the main components of the stone as well as its micro-components (crystalline phases), order of appearance, morphology, and location in the urinary tract. Techniques such as SEM/EDX and micro-XCT are becoming indispensable for studying the inorganic parts of kidney stones [5]. However, kidney stones are not just biominerals; they are heterogeneous biocomposites [6]. The development of a stone involves the formation of minerals within a slurry of substances (urine) that includes macromolecules, metabolites, vesicles, and other micro solids, in contact with cells. More research is needed on the organic and biological characterization of kidney stones, and on techniques that facilitate this task, particularly since biological components can exert catalytic roles on biomineralization. The study of kidney stones should be complemented with that of biofluids (blood, urine) and other factors that may be related to the type of renal calculi, such as associated pathologies, diet, and environmental factors.

It is important to hypothesize and explore potential links between kidney stones and variables across the diet, genome, microbiome, and exposome. The changes in the modern diet and environmental factors (exposome) should spur new hypotheses behind the increasing prevalence of idiopathic kidney stones [7]. Do specific dietary supplements increase or decrease the risk of kidney stone formation? How does stone prevalence map onto the quality of the water supply by geographical region? What about exposure to pollutants or microplastics? Differences in the urinary and intestinal microbiome between stone formers and controls have been reported, with oxalate-degrading bacteria receiving particular attention [8–10]. How does the presence or absence of microorganisms and their metabolites relate to kidney stones? Moreover, while researchers have identified a range of genetic factors linked to different types of kidney stone disease, the potential role of epigenetics remains largely unexplored [11].

Similarly, the connection between kidney stones and the broader human pathophysiology should be investigated. Advances in the understanding of lithiasis have revealed connections between this pathology and others like obesity, diabetes, renal tubular acidosis, osteoporosis, cardiovascular calcifications, epilepsy, etc. Some associations may be due to physicochemical conditions created by the pathology or the use of certain drugs, but they may still offer valuable mechanistic information. For example, the immune system is activated in the presence of certain types of stones [12]. Does it help dissolve stone precursors or may it aggravate growth? Urinary vesicles or exosomes are released across the urinary tract and capture rich information about the status of the kidney [13]. What can liquid biopsies tell us about kidney stones over time? Tools such as liquid biopsies, omics, bioimaging, microscopies, spectroscopies, and patient records offer high-dimensional datasets with potentially valuable information on urolithiasis. Artificial intelligence tools, which allow extracting patterns from high-dimensional data, may help develop novel risk scores, diagnostics, and therapeutics, among others. Clinical trials will be essential in assessing the value of new tools.

Basic research should not be limited to highly simplified systems, as this may take a very long time to advance past the initial stages of crystal nucleation or to discover principles relevant to patients. For example, many stones grow to form layers of organic and inorganic material, and it is often unclear whether the complex organic matrix accelerates or delays further mineral deposition. A simple experiment in which a stone is halved and incubated in concentrated urine may reveal which areas accrete minerals and which do not, providing valuable information despite the high complexity of the system. Urine transplants might enable testing hypotheses involving large numbers of interacting biochemical and biological species. Randall's plaque and urinary microcrystals are often associated with kidney stone disease, although they are also found in "healthy" individuals [9, 12]. Ingenious experiments could also be devised to assess the effects of different classes of potential stone precursors.

It is imperative that researchers and clinicians engage in ongoing education and training to keep pace with the constant advancement of knowledge. However, it is important to approach the hype surrounding certain technologies with caution and not allow it to dictate the direction of research or discourage those with limited resources from making valuable contributions. In fact, many low-cost technologies are underutilized, such as time-lapse microscopy, electron microscopy, isoelectric focusing, or Z-potential analysis, and may provide insight into important questions and even facilitate the development of novel clinical screening methods. It should also be noted that some national laboratories and synchrotron radiation facilities offer temporary access to equipment and expertise for scientists with sound hypotheses. To make progress in this field, it is necessary to consider multiple scales (temporal, spatial, and biological dimensions) and incorporate insights from various disciplines, including biochemistry, physical chemistry, materials science, geology, molecular biology, systems biology, physiology, and artificial intelligence.

In conclusion, there are many exciting research frontiers in the field of urolithiasis. Further understanding of the causes, mechanisms, and consequences of kidney stone formation will enable the development of more effective preventives, diagnostics, and therapeutics. This will be aided by the characterization of kidney stones and the corresponding fluids, as well as the exploration of links between kidney stones and variables across the diet, genome, microbiome, and exposome. Additionally, the connections between kidney stones and other pathologies, as well as the potential roles of the immune system and urinary vesicles, should be further investigated. Finally, the use of artificial intelligence and machine learning may have a role in improving the diagnosis and treatment of kidney stone disease.

**Acknowledgements**

The authors thank Emmanuel Letavernier (Sorbonne University), Oriol Angerri (Fundació Puigvert), and Pietro Manuel Ferraro (Agostino Gemelli University Policlinic) for valuable discussions.